\documentclass[reprint,nofootinbib,amsmath,amssymb,aps,floatfix]{revtex4-2}
\usepackage{graphicx}
\usepackage{dcolumn}
\usepackage[colorlinks=true,linkcolor=blue,urlcolor=magenta,filecolor=black,citecolor=magenta,pdfstartview=FitV,pdftitle={},pdfsubject={},pdfkeywords={},pdfpagemode=None,bookmarksopen=true]{hyperref}
\usepackage[multiple]{footmisc}
\usepackage{appendix}
\usepackage{color}
\usepackage{multirow}

\newcommand{\nn}{\nonumber}

\begin{document}
\title{Stability of five-dimensional Myers-Perry black holes under massive scalar perturbation: bound states and quasinormal modes}
\author{Wenbin Li\textsuperscript{1}}
\author{Kai-Peng Lu\textsuperscript{1}}
\author{W. LiMing\textsuperscript{1}}
\email{wliming@scnu.edu.cn}
\author{Jia-Hui Huang\textsuperscript{1,2,3}} 
\email{huangjh@m.scnu.edu.cn}
\date{\today}

\affiliation{$^{1}$ Key Laboratory of Atomic and Subatomic Structure and Quantum Control (Ministry of Education), School of Physics, South China Normal University, Guangzhou 510006, China}
\affiliation{$^{2}$ Guangdong Provincial Key Laboratory of Quantum Engineering and Quantum Materials, South China Normal University, Guangzhou 510006, China}
\affiliation{$^{3}$ Guangdong-Hong Kong Joint Laboratory of Quantum Matter, Frontier Research Institute for Physics, South China Normal University, Guangzhou 510006, China}

\begin{abstract}
The stability of five-dimensional singly rotating Myers-Perry Black Holes against massive scalar perturbations is studied. 
Both the quasibound states and quasinormal modes of the massive scalar field are considered. For the quasibound states, we use an analytical method to 
discuss the effective potential felt by the scalar field, and found that there is no potential well outside the event horizon. Thus, 
singly rotating Myers-Perry Black Holes are stable against the perturbation of quasibound states of massive scalar fields.  
Then, We use continued fraction method based on solving a seven-term recurrence relations to compute the spectra of the quasinormal modes.
For different values of the black hole rotation parameter $a$, scalar mass parameter $\mu$ and angular quantum numbers, all found quasinormal modes
are damped. So singly rotating Myers-Perry Black Holes are also stable against the perturbation of quasinormal modes of massive scalar fields. 
Besides, when the scalar mass $\mu$ becomes relatively large, the long-living quasiresonances are also found as in other rotating black hole models. 
Our results complement previous arguments on the stability of five-dimensional singly rotating Myers-Perry black holes against massive scalar perturbations. 
\end{abstract}
\maketitle

\section{Introduction}\label{sec.1}

The history of black holes (BHs) is quite long. The BH physics plays an essential role in modern theoretical physics and observational physics. 
According to the BH perturbation theory, when a BH is perturbed by any possible fields (scalar, electromagnetic, or gravitational), these perturbations are usually governed by a pair of second-order ordinary differential equations.
The physical requirement of the perturbation fields must be purely ingoing at the event horizon and are all finite near spatial infinity leads to three different boundary conditions.
They are respectively called scattering states, quasinormal modes (QNMs) and quasibound states (QBSs). 
The well-known QNMs play important roles during the ringdown process.
They are independent of the initial conditions, reflect the intrinsic properties of the spacetime itself,
and are regarded as the characteristic modes of the oscillations of BHs \cite{Berti:2009kk,Konoplya:2011qq}.

It was suggested that particles with negative energy can exist in the ergoregion of a rotating BH, 
so one can imagine a process through which it may be possible to extract energy from a rotating BH, which is the Penrose’s process
\cite{Penrose:1971uk,Christodoulou:1970wf}. 
Similarly, there exists an analog effect for an incident bosonic wave, 
which is called superradiance \cite{Zeldovich:1972,Misner:1972kx,Bardeen:1972fi,Bekenstein:1973mi,Press:1972zz,Teukolsky:1974yv}.
It is said that when a bosonic wave is impinging upon a rotating BH, the reflected wave will be amplified by 
extract the rotational energy from BHs if the frequency $\omega$ of the wave satisfies $\omega < m \Omega_H$, where $m$ is the azimuthal number with respect to the BH rotation axis and $\Omega_H$ represents the angular velocity of the BH event horizon. 
(For a comprehensive review about superradiance, see \cite{Brito:2015oca}.)
It was pointed out that when putting an artificial mirror outside the event horizon, 
the amplified waves may be reflected back and forth, thus leading to an instability. This is 
the so-called ``black hole bomb'' mechanism \cite{Press:1972zz,Cardoso:2004nk}.
This instability was later realized by imposing (charged) massive scalar perturbations on Kerr (or Kerr-Newman) BH background \cite{Detweiler:1980uk,Furuhashi:2004jk,Konoplya:2006br,Dolan:2007mj}. In these cases, the mass term of the perturbation field behaves as a natural mirror. Meanwhile, the superradiant (in)stability of asymptotically flat Kerr BHs under massive scalar or vector perturbations has been studied in Ref.\cite{Cardoso:2011xi,Hod:2012zza,Dolan:2012yt,Hod:2016bas,East:2017ovw,Huang:2018qdl,Huang:2019xbu}.

Higher-dimensional spacetimes are also interesting in theoretical physics. On the one hand, higher dimensions are necessary for string theory, compact extra dimensions, brane-world models, gauge/gravity duality, etc. On the other hand, in four-dimensional general relativity, a uniqueness theorem which was proved by Carter and Robinson states that the only possible stationary and axial-symmetric flat spacetime is the Kerr solution \cite{Carter:1971zc,Robinson:1975bv}. But it does not hold in higher-dimensional spacetime, where there exist a variety of black object solutions 
such as black strings, branes, rings and so on \cite{Emparan:2008eg}.

Compared to four-dimensional cases, the (in)stability of higher-dimensional BHs are more complicated. 
The higher-dimensional spherically symmetric Schwarzschild-Tangherlini spacetime and Reissner-Nordstr\"{o}m (RN) BHs were proven to be stable when perturbed by external fields\cite{Molina:2003ff,Ishibashi:2003ap,Konoplya:2008rq,Huang:2021jaz,Huang:2022nzm,Huang:2021,Huang:2022}. However, the black strings and p-branes constructed from the spherically symmetric BHs are generally unstable, this is known as the Gregory–Laflamme instability \cite{Gregory:1993vy}. 
The superradiant conditions of higher-dimensional rotating BHs was studied in Ref.\cite{Jung:2005cn,Jung:2005nf}. 
It was shown that Myers-Perry (MP) BHs under AdS background with equal angular momenta in odd number of dimensions (greater than five) are superradiant unstable under tensor perturbations \cite{Kunduri:2006qa}. The scalar QNMs were numerically explored for $Kerr-AdS_5$ with unequal rotations \cite{Koga:2022vun}. Superradiance also leads to gravitational instability
in other four and higher-dimensional AdS BHs \cite{Cardoso:2006wa,Kodama:2007sf,Murata:2008xr,Cardoso:2013pza}. 
It has been recently proven that small MP-AdS BHs in arbitrary dimensions are also superradiant unstable \cite{Delice:2015zga}. In Ref.\cite{Gwak:2019ttv}, the author has derived a near-extremal QNMs formula of higher-dimensional singly rotating MP-dS BHs with non-minimally coupled scalar fields, and singly rotating MP-dS black hole was found to be stable against the QNMs of massive scalar perturbations\cite{Ponglertsakul:2020ufm}.

Asymptotically flat Myers-Perry Black Holes (MPBHs) with equal angular momenta are found to be stable under gravitational perturbation in five or seven dimensions,
but in nine dimensions, for sufficiently rapid rotation, the authors found that perturbations grow exponentially in time \cite{Dias:2010eu}. As for $D \geq 7$, the stability of simply rotating MPBHs against tensor-type perturbations was studied and no 
instability was found \cite{Kodama:2009bf}. Moreover, the QNMs of massless scalar perturbations on $D=5$ and $D=6$ MP spacetime with a single rotation parameter were well investigated,
and the numerical results shown that they are all stable \cite{Ida:2002zk,Cardoso:2004cj}. 
The QNMs of massive scalar perturbations on $D=6$ MPBH with a single rotation parameter has been studied recently and no instability is found\cite{Lu:2023par} . 

In Ref.\cite{Cardoso:2005vk}, the authors considered the stability of singly rotating MPBHs against massive scalar perturbations.
Based on the assumption that there are no stable orbits for $D>4$ MPBHs, and thus the perturbation field can escape to infinity, they argued that 
singly rotating MPBHs should be stable against massive scalar perturbations.
In order to provide a direct and complementary evidence for the stability of a $5D$ singly rotating MPBH against massive scalar perturbations,
in this work, we consider the stability of the MPBH against both the QBSs and QNMs of massive scalar perturbations \cite{Rosa:2011my}.
We adopt an analytical method to discuss the QBSs and use the continued fraction method to study the QNMs.

This paper is organized as follows. In Sec.\ref{sec.2}, we briefly review the higher-dimensional singly rotating MPBH solution 
and the Klein-Gordon equation in a $5D$ singly rotating MPBH. From the radial EOM, we obtain the effective potential and discuss the
relevant boundary conditions. In Sec.\ref{sec.3}, we consider the asymptotic behavior of the effective potential, then analytically study the effective potential and discuss the stability of $5D$ MPBH against the QBSs of the massive scalar fiels. 
In Sec.\ref{sec:4}, we first introduce the continued fraction method used here, and then compute and discuss the QNMs of the massive scalar 
fields for different values of rotation parameter and scalar mass. The final section is devoted to the conclusions.

\section{The Background Metric and effective potential}\label{sec.2}

A higher-dimensional generalization of the Kerr metric to an arbitrary number of dimensions $D \geq 5$ was discovered by Myers and Perry in Ref.\cite{Myers:1986un}.For a ($4+n$)-dimensional MPBH, there are $\lfloor\dfrac{n+3}{2}\rfloor$ independent angular momentum components, each of which corresponds to a rotation plane. Throughout this work, we concentrate on the simplest case  where the MPBH rotates in just one plane, and we denote the angular momentum (per mass) of the MPBH by $a$. Thus, the line element of a singly rotating higher-dimensional MPBH in Boyer-Lindquist-type coordinates is given by

\begin{align}
	\mathrm{d} s^2 =& - \dfrac{\Delta_n - a^2 \sin^2\theta}{\Sigma} \mathrm{d} t^2 - \dfrac{2 a M r^{1-n} \sin^2\theta}{\Sigma} \mathrm{d} t \mathrm{d} \varphi \nn \\
	&+ \dfrac{(r^2+a^2)^2 - \Delta_n a^2 \sin^2 \theta }{\Sigma} \sin^2 \theta \mathrm{d} \varphi^2 + \dfrac{\Delta_n}{\Sigma} \mathrm{d} r^2 \nn \\
	&+ \Sigma \mathrm{d} \theta^2 + r^2 \cos^2 \theta \mathrm{d} \Omega^2_n, \label{2.1}
\end{align}
where
\begin{align}
	\Sigma &= r^2 + a^2 \cos^2 \theta, \label{2.2} \\
	\Delta_n &= r^2 + a^2 - M r^{1-n}. \label{2.3}
\end{align}
$\mathrm{d} \Omega^2_n$ denotes the standard metric of the unit $n$-sphere. $M,a$ are related to the physical mass $M_{\rm BH}$ and angular momentum $J$ of the MPBH as follows\cite{Berti:2003yr},
\begin{equation}
	M_{\rm BH} = \dfrac{(n+2) \mathcal{A}_{n+2} M }{16 \pi G},\quad J = \dfrac{\mathcal{A}_{n+2} M a}{8 \pi G},
\end{equation}
where $\mathcal{A}_{n+2} = 2 \pi ^{\frac{n+3}{2}} / \Gamma\left( \frac{n+3}{2} \right)$ represents the area of a unit $(n+2)$-dimensional sphere and $G$ is the ($4+n$)-dimensional Newtonian constant of gravitation. The above metric describes an asymptotically flat and rotating vacuum BH solution with spherical topology. Hereafter, without loss of generality, $M,a > 0$ are assumed.

The event horizon of the BH is located at $r=r_H$, which is the largest real root of the equation $\left. \Delta_n \right|_{r=r_H} = 0$. For $n=0$, the above metric is just the $4D$ Kerr BH, and we are familiar with the two event horizons for a Kerr BH. Note that there is only one horizon when $n \geq 1$. For $n=1$, which is the case we focus on in this paper, the rotation parameter is bounded above, $a < \sqrt{M}$.

According to the BH perturbation theory, the dynamical evolution of a massive scalar perturbation field $\Psi(x)$ with mass $\mu$ in the background spacetime \eqref{2.1} is governed by the covariant KG equation
\begin{equation}\label{2.5}
	\Box \Psi(x) = \dfrac{1}{\sqrt{-g}} \partial_\alpha \left[ g^{\alpha\beta} \sqrt{-g} \partial_\beta \Psi(x) \right]  = \mu^2 \Psi(x), 
\end{equation}
where $g = \det(g_{\alpha\beta})$ is the determinant of the spacetime metric. Since Eq.\eqref{2.5} is separable in these Boyer-Lindquist-type coordinates $x^\mu = \left\lbrace t,r,\theta,\varphi,\cdots \right\rbrace $, so we can decompose the eigenfunction with the following ansatz \cite{Ida:2002zk,Cardoso:2004cj,Cardoso:2005vk}
\begin{equation}\label{2.6}
	\Psi(x) = e^{-i \omega t} e^{i m \varphi} R(r) S(\theta) Y(\Omega),
\end{equation}
where $Y(\Omega)$ is the hyperspherical harmonics on the $n$-sphere with eigenvalues $-j(j+n-1)(j=0,1,2,\cdots)$. The quantum number $m(=0,\pm 1,\pm 2,\cdots)$ describes the dependence of the perturbation field on the azimuthal direction around the rotation axis.  Substitute the ansatz into Eq.\eqref{2.5}, the scalar $(4+n)$-dimensional spheroidal harmonics $S(\theta)$ satisfies the following angular EOM
\begin{align}\label{2.7}
	\dfrac{1}{\sin \theta \cos^n \theta} \bigg( \dfrac{\mathrm{d}}{\mathrm{d}\theta} & \sin \theta \cos^n \theta \dfrac{\mathrm{d} S(\theta)}{\mathrm{d}\theta} \bigg) + \bigg[ c^2 \cos^2 \theta - \dfrac{m^2}{\sin^2 \theta} \nn \\ 
	&- \dfrac{j(j+n-1)}{\cos^2 \theta} + \lambda_{kjm} \bigg] S(\theta) = 0,
\end{align}
where $c = a \sqrt{\omega^2 - \mu^2}$. Different from the four-dimensional Kerr BH case, the angular separation constant $\lambda_{kjm}$ now depends on three indices $k,m,j$. The parameter $k(=0,1,2,\cdots)$ labels the discrete eigenvalues of $S(\theta)$ for fixed values of $j$ and $m$. When $a \omega \ll 1$ and $a \mu  \ll 1$, the separation constant $\lambda_{kjm}$ can be expanded as a Taylor series as follows
\begin{equation}\label{2.8}
	\lambda_{kjm} = \ell (\ell+ n + 1) + \sum_{p = 1}^{\infty} \tilde{f}_p c^p,
\end{equation}
where $\ell= 2k + j + |m|$ is an integer which satisfies $\ell \geq j + |m|$, and the first five terms of $\tilde{f}_p$ are shown explicitly in Ref.\cite{Berti:2005gp}. 

%\section{The Radial EOM and Boundary Conditions}\label{sec.3}

From the KG equation \eqref{2.5}, we can also obtain the following radial EOM which is obeyed by $R(r)$,
\begin{equation}\label{3.1}
	\dfrac{1}{r^n} \dfrac{\mathrm{d}}{\mathrm{d} r} \left( r^n \Delta_n \dfrac{\mathrm{d} R(r)}{\mathrm{d} r} \right) + U(r) R(r) = 0,
\end{equation}
where
\begin{widetext}
	\begin{equation}\label{3.2}
		\begin{aligned}
			U(r) = \dfrac{\left[ \omega ( r^2 + a^2 ) - ma \right]^2}{\Delta_n} - \bigg[ \dfrac{j(j+n-1) a^2}{r^2} + \lambda_{kjm} - 2 m a \omega + \mu^2 r^2 + a^2 \omega^2 \bigg].	
		\end{aligned}
	\end{equation}
\end{widetext}
It's easy to check that the above radial EOM  may reduce to the four-dimensional Teukolsky equation \cite{Press:1972zz,Bardeen:1972fi,Teukolsky:1974yv} when $n=0$ (hence $M = 2 G M_{\rm BH}$). 

In order to study the appropriate boundary conditions of the scalar perturbation, it is useful to define a tortoise coordinate by $\dfrac{\mathrm{d} r_*}{\mathrm{d} r} = \dfrac{r^2 + a^2}{\Delta_n}$ and a new radial function $\tilde{R}(r) = \sqrt{r^n (r^2 + a^2)} R(r)$. With these new quantities, the radial EOM is transformed into the following Schr\"{o}dinger-like equation
\begin{equation}\label{3.3}
	\dfrac{\mathrm{d}^2 \tilde{R}(r)}{\mathrm{d} r_*^2} + \tilde U(r) \tilde{R}(r) = 0.
\end{equation}
Then, the asymptotic limits of $\tilde U(r)$ at the spatial infinity and event horizon are 
\begin{align}\label{3.4}
	\tilde U(r) \sim 
	\begin{cases}
		\left( \omega - m \Omega_H \right)^2, & \quad {} \, r_* \rightarrow -\infty \, (r \rightarrow r_H), \\
		\omega^2 - \mu^2, & \quad {} \, r_* \rightarrow +\infty \, (r \rightarrow +\infty),
	\end{cases}
\end{align}
where $\Omega_H = \dfrac{a}{r_H^2 + a^2}$ is the angular velocity of the BH event horizon. The physically accepted boundary condition of the perturbation field at the classical BH horizon is a purely ingoing wave. Then, the 
asymptotic solutions of the radial equation \eqref{3.3} at the horizon and spatial infinity are chosen as follows,
\begin{equation}\label{3.5}
	R(r) \sim 
	\begin{cases}
		(r-r_H)^{- i \sigma}, & \quad  r \rightarrow r_H, \\
		r^{-(n+2)/2} {\rm e}^{q r}, & \quad  r \rightarrow \infty,
	\end{cases}
\end{equation}
where
\begin{equation}
	q^2 = \mu^2-\omega^2,\quad \sigma = \dfrac{\left[ \left( r_H^2 + a^2 \right) \omega - m a \right] r_H}{ (n-1) \left( r_H^2 + a^2 \right) + 2 r_H^2}.
\end{equation}
At spatial infinity, two physical boundary conditions may exist. One is the renowned QNM condition which imposes purely outgoing modes at $r \rightarrow \infty$. In this case, 
${\rm Re}(\omega)>\mu$ and $q=i\sqrt{\omega^2-\mu^2}$. The other is the quasibound state condition which requires decaying modes at $r \rightarrow \infty$. In this case,   
${\rm Re}(\omega)<\mu$ and $q=-\sqrt{\mu^2-\omega^2}$.  

The radial equation \eqref{3.1} and  chosen boundary conditions single out a discrete set of complex frequencies $\left\lbrace \omega_n \right\rbrace$ ( $\omega_n \equiv \omega_R - i \omega_I$). In our convention,  $\omega_I > 0$ means that the perturbation field is a decaying and stable mode, while $\omega_I < 0$ implies an growing and instable mode \cite{Berti:2009kk,Konoplya:2011qq}.

\section{Stability analysis of the QBSs}\label{sec.3}

In this section, we consider the bound state condition and analytically study the superradiant stability regime of the $5D$ rotating MPBH under massive scalar perturbations. The event horizon of the MPBH is the solution of $\Delta_1 = 0$, i.e. $r_H = \sqrt{M - a^2}$. For bound state, the radial function is decaying at spatial infinity, i.e. $q=-\sqrt{\mu^2-\omega^2}$ in the asymptotic solution \eqref{3.5}. In order for a superradiant scattering occurring, the angular frequency $\omega$ should satisfy the following inequality
\begin{equation}\label{4.1}
	\omega < \omega_c \equiv m \Omega_H = \dfrac{m a}{M}.
\end{equation}
Since the massive scalar field is considered as a perturbation of the MPBH, it is required that the mass of the scalar field is in fact much less than the mass of the MPBH. 
In our $5D$ MPBH case, this requirement means $\mu^2 M\ll 1$.

\subsection{Asymptotic Analysis of the effective potential}\label{sec:4.1}

According to equation \eqref{3.1},  the radial EOM of a scalar field in $5D$ MPBH ($n=1$) can be written as
\begin{equation}\label{4.2}
	\Delta_1 \dfrac{1}{r} \dfrac{\mathrm{d}}{\mathrm{d} r} \left( r \Delta_1 \dfrac{\mathrm{d} R(r)}{\mathrm{d} r} \right) + W(r) R(r) = 0,
\end{equation}
where 
\begin{align}
	W(r) =& \left[ \omega ( r^2 + a^2 ) - m a \right]^2 \nn \\
	& - \Delta_1 \bigg( \dfrac{j^2 a^2}{r^2} + \lambda_{kjm} - 2ma \omega + \mu^2 r^2 \bigg),\label{4.3}
\end{align}
Defining a new function  $\psi(r) = \sqrt{r \Delta_1} R(r)$ \cite{Hod:2012zza}, the radial EOM can be rewritten as 
\begin{equation}\label{4.4}
	\dfrac{\mathrm{d}^2 \psi(r)}{\mathrm{d} r^2} + \left[ \omega^2 - V_{\rm eff}(r) \right] \psi(r) = 0,
\end{equation}
where the effective potential is 
\begin{equation}\label{4.5}
	V_{\rm eff}(r) = \omega^2 + \dfrac{- 4 r^2 W(r) - \Delta_1 ^2 + 4 r^2 \left( 2 \Delta_1 - r^2 \right)}{4 r^2 \Delta_1^2}.
\end{equation}

The superradiant (in)stability can be judged by analyzing whether the effective potential has a potential well outside the event horizon \cite{Hod:2012zza,Cardoso:2005vk}. 
If there was a potential well, the superradiant QBS may be trapped and scattered back and forth (black hole ``bomb" mechanism\cite{Press:1972zz,Cardoso:2004nk}), which leads to the superradiant instability of the system. Here it is worth emphasizing that the existence of a trapping potential well is a necessary condition (but not a sufficient one) for the instability of the system. On the other hand, if there was
no potential well, one could conclude that the system is superradiantly stable. 

Note that in principle it's not enough to analyze the existence of the potential well by just considering the asymptotic behaviour of the effective potential at spatial infinity. As discussed in Ref.\cite{Mai:2021yny}, one may find a potential well sandwiched between two barriers. So one needs to analyze the existence of the potential well in the whole spatial region outside 
the event horizon. 

We first consider the asymptotic behaviors of the effective potential \eqref{4.5} and its derivative at the horizon and spatial infinity, which are
\begin{widetext}
	\begin{align}
		V_{\rm eff} ( r \rightarrow r_H ) &\longrightarrow - \infty, \quad V_{\rm eff}' (r \rightarrow r_H ) \longrightarrow + \infty, \label{4.6} \\
		V_{\rm eff} (r \rightarrow +\infty ) &\longrightarrow \mu ^2 + \frac{M \left( \mu ^2 - 2 \omega ^2 \right)-a^2\mu^2 + \lambda_{kjm} + \dfrac{3}{4}}{r^2} + \mathcal{O} \left(\dfrac{1}{r^3}\right), \label{4.7} \\
		V_{\rm eff}' (r \rightarrow +\infty ) &\longrightarrow \frac{- 2 M \left( \mu ^2 - 2 \omega ^2 \right) +2a^2\mu^2 -2 \lambda_{kjm}- \dfrac{3}{2}}{r^3} + \mathcal{O} \left(\dfrac{1}{r^4}\right). \label{4.8}
	\end{align}
\end{widetext}
It is easy to see that the effective potential approaches to a constant $\mu^2$ at spatial infinity. 
As we mentioned below equation \eqref{4.1}, perturbation analysis requires that $\mu^2 M\ll 1$. For bound state, $\omega^2<\mu^2$, thus $\omega^2 M\ll 1$.
Since $a^2<M$, we have $a^2 \mu^2\ll 1$. Under these conditions,  $\lambda_{kjm}\simeq \ell (\ell+ n + 1)>0 $. 
Now it's easy to see the numerator of the leading order of $V_{\rm eff}'$ is negative, i.e. 
$- 2 M \left( \mu ^2 - 2 \omega ^2 \right) +2a^2\mu^2 -2 \lambda_{kjm}- \dfrac{3}{2}<0$.

Based on the above analysis, we find that $V_{\rm eff}' \rightarrow 0^-$ as $r \rightarrow \infty$, i.e. there is no trapping well for the effective potential near spatial infinity.
Then, according to the asymptotic behaviours of the effective potential at the event horizon and spatial infinity, we can infer that there is at least one maximum between the event horizon and spatial infinity, i.e. there is at least one potential barrier between the event horizon and spatial infinity.
However, as discussed before, this is not sufficient to conclude that the system is superradiantly stable. We need a further analysis on the shape of the effective potential 
between the event horizon and the spatial infinity.  

In the next subsection, we use an analytic method based on Descartes' rule of signs to show that there is no trapping well for the effective potential outside the event horizon.

\subsection{Analysis of the Derivative of the Effective Potential}\label{sec:4.2}

In this subsection, we analyze the shape of the effective potential by considering the positive real roots of the algebraic equation $V_{\rm eff}'(r) = 0$ in the physically allowed interval $(r_H,+\infty)$. The derivative of the effective potential is 
\begin{align}\label{4.10}
	V_{\rm eff}'(r) = - \dfrac{f(r)}{2 r^3 \Delta_1^3} ,
\end{align}
where 
\begin{align}
	f(r) =& A_1 r^6 + B_1 r^5 + C_1 r^4 + D_1 r^3 + E_1 r^2 + F_1 r + G_1 \nn \\
	=& \left[ 4 M \left( \mu ^2-2 \omega ^2 \right) + 4 \lambda_{kjm} + 3 \right] r^6 \nn \\
	&+ \left[ 16 a m M \omega - M \left( 4 M \mu ^2 + 4 \lambda_{kjm} + 9 \right) \right. \nn \\
	&\qquad \left. + a^2 \left( 8 j^2 - 8 m^2 + 4 \lambda_{kjm} + 9 \right) \right] r^4 \nn \\
	&- \left\lbrace 3 r_H^2 \left[ a^2 \left( 4 j^2 - 1 \right) + M \right] \right\rbrace r^2 \nn \\
	&+ r_H^4 \left[ a^2 \left( 4 j^2 - 1 \right) + M \right].
\end{align}
\begin{widetext}
	Because we are interested in the real roots of $V_{\rm eff}'(r) = 0$ when $r > r_H$, we can ignore the nonzero denominator of $V_{\rm eff}'(r)$
	and equivalently consider the real roots of $f(r)=0$.
	Defining a new variable, $z \equiv r - r_H$, $f(r)$ can be rewritten as
	\begin{align}\label{4.10}
		f(r) = f_1(z) = A z^6 + B z^5 + C z^4 + D z^3 + E z^2 + F z + G,
	\end{align}
	where
	\begin{align}
		A =& \, 4 M \left( \mu ^2-2 \omega ^2 \right) + 4 \lambda_{kjm} + 3, \\
		B =& \, 6 r_H \left[ 4 M \left( \mu ^2-2 \omega ^2 \right) + 4 \lambda_{kjm} + 3 \right], \\
		C =& \, 16 a m M \omega + 4 M \left[ 14 M \left( \mu ^2 - 2 \omega ^2 \right) - 2M \omega^2 + 14 \lambda_{kjm} + 9 \right] \nn \\
		&- a^2 \left( - 2 j^2 + 2 m^2 + 14 \lambda_{kjm} + 9 \right) , \\
		D =& \, 8 r_H \left\lbrace 8 a m M \omega + M \left[ 8 M \left( \mu ^2  - 2 \omega ^2 \right) - 4 M \omega^2 + 8 \lambda_{kjm} + 3 \right] \right. \nn \\
		&- \left. a^2 \left( - 4 j^2 + 4 m^2 + 8 \lambda_{kjm} + 3 \right) \right\rbrace , \\
		E =& \, 12 r^2_H \left\lbrace 8 a m M \omega + M \left[ 3 M ( \mu ^2 - 2 \omega ^2 )- 4 M \omega^2 + 3 \lambda_{kjm} - 1 \right] \right. \nn \\
		&- \left. a^2 \left( - 3 j^2 + 4 m^2 + 3 \lambda_{kjm} - 1 \right) \right\rbrace, \\
		F =& \, 8 r_H^3 \left\lbrace 8 a m M \omega + M \left[ M \left(\mu ^2 - 6 \omega ^2 \right) + \lambda_{kjm} - 3 \right] \right. \nn \\
		&- \left. a^2 \left[ - j^2 + 4 m^2 + \lambda_{kjm} - 3 \right] \right\rbrace , \\
		G =& -8 r_H^4 \left[ - 2 a m M \omega + M \left( M \omega^2 + 1 \right) + a^2 \left( m^2 - 1 \right) \right].
	\end{align}
\end{widetext}
The real roots of $f(r) = 0$ with $r > r_H$ are one-to-one corresponding to the positive real roots of $f_1(z)=0$ with $z>0$. 

In order to use the method based on the Descartes' rule of signs, we analyze the signs or sign relations of the coefficients in $f_1(z)$. 
First, given the inequalities $\omega^2 M\ll 1, \mu^2 M\ll 1$, it is easy to get the following results 
\begin{equation}
	A > 0, \quad B > 0.
\end{equation}
Then, considering $G$ as a quadratic function of $\omega$, it is easy to see that it opens downward with discriminant $\Delta_G = - 256 M^2 r_H ^{10} < 0$. Therefore, we have
\begin{equation}
	G < 0.
\end{equation}

It is not easy to directly judge the signs of the other coefficients. In the next, we study the sign relations between pairs of adjacent coefficients. We introduce the following scaled new coefficients 
\begin{equation}\label{4.20}
	C' = \dfrac{C}{2},\quad D' = \dfrac{D}{8 r_H},\quad E' = \dfrac{E}{12 r_H^2},\quad F' = \dfrac{F}{8 r_H^3}.
\end{equation}
It is worth noting that all scaling factors in the above equations are positive. Therefore, $C$ and $C'$ ( $D$ and $D',\cdots$) possess the same sign, i.e. they are simultaneously positive or negative. Taking the difference between $C'$ and $D'$, given the inequalities $\omega^2 M\ll 1, \mu^2 M\ll 1$, we can obtain
\begin{align}\label{4.22}
	C' - D' =& \dfrac{C}{2} - \dfrac{D}{8 r_H} \nn \\
	=& 5 r_H^2 \left[4 M \left( \mu^2 - 2 \omega^2 \right) + 4 \lambda_{kjm} + 3 \right] > 0.
\end{align}
Next, let’s calculate the difference between $D'$ and $E'$,
\begin{align}
	D' - E' =& \dfrac{D}{8 r_H} - \dfrac{E}{12 r_H^2} \nn \\ 
	=& M \left[ 5 M \left(\mu ^2-2 \omega^2 \right) + 5 \lambda_{kjm} + 4 \right] \nn \\
	&- a^2 \left( - j^2 + 5 \lambda_{kjm} + 4 \right).
\end{align}
Given the inequalities $\omega^2 M\ll 1, \mu^2 M\ll 1$, it's obvious that the coefficient of $M$ is greater than the coefficient of $a^2$. Together with the inequality $M>a^2$, 
so we have
\begin{equation}\label{4.24}
	D' > E'.
\end{equation}
Similarly, we can also obtain
\begin{align}\label{4.25}
	E' - F' =& \dfrac{E}{12 r_H^2} - \dfrac{F}{8 r_H^3} \nn \\
	=& 2 \big\lbrace M \big[ M \big(\mu ^2 - 2 \omega^2 \big) + \lambda_{kjm} + 1 \big] \nn \\
	&- a^2 \big( - j^2 + \lambda_{kjm} + 1 \big) \big\rbrace  > 0,
\end{align}
According to the above three inequalities \eqref{4.22},\eqref{4.24} and \eqref{4.25}, we finally obtain
\begin{equation}\label{4.26}
	C' > D' > E' > F'.
\end{equation}

%Table I
\begin{table}[]
	\centering
	\caption{All possible signs of the coefficients ($A, B, C,\dots, G$).}
	\label{tab:1}
	\renewcommand\arraystretch{1.5}
	\setlength{\tabcolsep}{4mm}{
		\begin{tabular}{c|c|c|c|c|c|c}
			\hline % horizontal line
			\hline
			$A$ & $B$ & $C$ & $D$ & $E$ & $F$ & $G$ \\
			\hline
			\multirow{5}{*}{$+$} & \multirow{5}{*}{$+$} & \multirow{4}{*}{$+$} & \multirow{3}{*}{$+$} & \multirow{2}{*}{$+$} & + & \multirow{5}{*}{$-$} \\
			\cline{6-6}
			&  &  &  &  & $-$ &  \\
			\cline{5-6}
			&  &  & & $-$ & $-$ &  \\
			\cline{4-6}
			&  &  & $-$ & $-$ & $-$ &  \\
			\cline{3-6}
			&  & $-$ & $-$ & $-$ & $-$ &  \\
			\hline
			\hline
	\end{tabular}}
\end{table}

In mathematics, Descartes' rule of signs provides a practical theorem to determine the possible number of positive real roots of a polynomial equation. It says that if a polynomial with real coefficients is arranged in descending order of powers, then the number of positive real roots of the polynomial is either equal to the number of sign changes between adjacent non-zero coefficients, or is less than it by an even number. For the polynomial equation $f_1(z)=0$, we show all possible signs of its coefficients in Table \ref{tab:1}. It is obvious that the number of sign changes of the coefficients $\left( A,B,C,D,E,F,G \right)$ is always $1$. 
So there is at most one positive real root for $f_1(z) = 0$, equivalently, for $f(r) = 0$ with $r>r_H$. And according to the previous asymptotic analysis, we know that there is at least one maximum for the effective potential outside the event horizon. 

\begin{figure}[]
	\centering
	\includegraphics[scale=0.65]{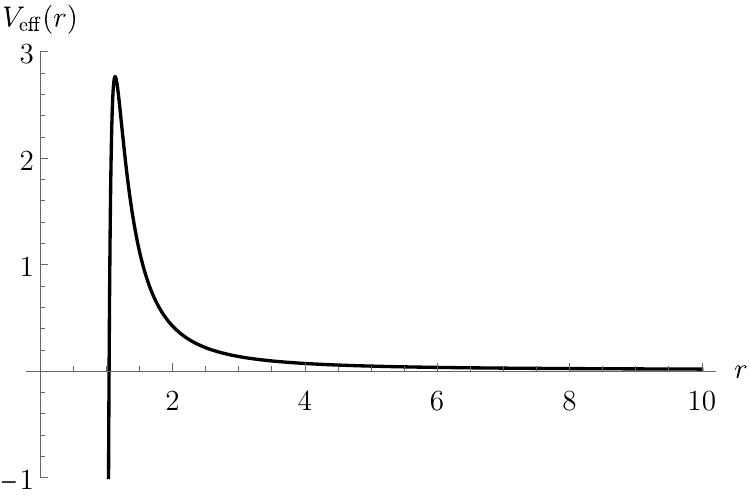}
	\caption{Typical shape of the effective potential in the radial EOM. The parameters are chosen as $M=1,a=0.4,k=j=0,m=1,\omega=0.1,\mu=0.2$.}
	\label{fig:1}
\end{figure}

We conclude that there is only one maximum point (potential barrier) outside the horizon $r_H$ for the effective potential. As an illustration, we show a typical effective potential in Fig.\ref{fig:1}. In this situation, the black hole "bomb" mechanism is not triggered, and the system consisting of MPBH and massive scalar perturbation is superradiantly stable.

%Figure 1

\section{Numerical Analysis about QNMs}\label{sec:4}

In this section, we consider the QNMs boundary condition, i.e. $q = i \sqrt{\omega^2 - \mu^2}$ in Eq.\eqref{3.5}, and numerically study the QNMs of the system consisting of MPBH and massive scalar perturbation. As mentioned in the reviews \cite{Berti:2009kk,Konoplya:2011qq}, there are a series of methods to calculate QNMs. Here we adopt  Leaver's continued fraction method \cite{Leaver:1985ax}, which is the most accurate technique to date to compute quasinormal frequencies. For simplicity, we set  $M = 1$ in this section, so that $r$ and $a$ are measured in units of $M^{1/2}$, while $\omega$ and $\mu$ are scaled with $M^{-1/2}$.

\subsection{Continued Fraction Method}\label{sec:5.1}

Defining a new variable, $u \equiv \cos \theta$, the $5D$ angular EOM \eqref{2.7}  can be written as follows
\begin{align}
	\dfrac{1}{u} \bigg[ \dfrac{\partial}{\partial u} & u \left( 1 - u^2 \right) \dfrac{\partial}{\partial u} \bigg] S(\theta) \nn \\
	&+ \bigg[ c^2 u^2 + \lambda_{kjm} - \dfrac{m^2}{1-u^2} - \dfrac{j^2}{u^2} \bigg] S(\theta) = 0.\label{5.1}
\end{align}
The angular function $S(\theta)$ can be assumed to have a series expansion form,
\begin{equation}\label{5.2}
	S = (1 - u^2)^{\frac{|m|}{2}} u^j \sum_{k = 0}^{\infty} a_k u^{2 k}.
\end{equation}
This series (if convergent) automatically satisfies regular boundary conditions at three singular points $\theta=0,\pi/2,\pi$ \cite{Berti:2005gp}. 

Substituting Eq.\eqref{5.2} into Eq.\eqref{5.1}, we obtain a three-term recurrence relations 
\begin{align}
	& \alpha_0^\theta a_1 + \beta_0^\theta a_0 =0, \nonumber \\
	& \alpha_k^\theta a_{ k+1 } + \beta_k^\theta a_k + \gamma_k^\theta a_{ k-1 } =0, \qquad ( k = 2, 3, \cdots),
\end{align}
where the superscript $\theta$ indicates that these symbols are used to describe the recurrence coefficients of the angular equation. The explicit forms of these coefficients are given by
\begin{align}
	\alpha_k^\theta &= -4 (1 + k) (j + k + 1), \nn \\
	\beta_k^\theta &= \left( 2k + j + |m| \right) \left( 2k + j + |m| + 2 \right) - \lambda_{kjm}, \nn \\
	\gamma_k^\theta &= -c^2.
\end{align}

Then the continued fraction equation for the separation constant $\lambda_{kjm}$  has the same form as in the $4D$ Kerr case \cite{Leaver:1985ax},
\begin{equation}\label{5.5}
	\beta_0^\theta - \frac{ \alpha_0^\theta \gamma_1^\theta}{ \beta_1^\theta - } \frac{ \alpha_1^\theta \gamma_2^\theta }{ \beta_2^\theta - } \frac{ \alpha_2^\theta \gamma_3^\theta }{ \beta_3^\theta -  } \cdots 
	\equiv
	\beta_0^\theta - \frac{\alpha_0^\theta \gamma_1^\theta}{\beta_1^\theta - \frac{\alpha_1^\theta \gamma_2^\theta} {\beta_2^\theta - \frac{\alpha_2^\theta \gamma_3^\theta}
			{\beta_3^\theta - \cdots}}}
	= 0.
\end{equation}
As $a \rightarrow 0$, the MPBH spacetime reduces to $5D$ Schwarzschild BH. All $\gamma_k^\theta$ will become zero, and the eigenvalue $\lambda_{kjm}$ can be calculated analytically, which is 
\begin{align}
	\lambda_{kjm} =& \left( 2 k + j + |m| \right) \left( 2k + j + |m| + 2 \right), \nn \\
	&(k=0,1,2,\cdots).\label{5.6}
\end{align}

Similarly, the solution of the radial EOM can also be expressed as a series expansion of the following form  
\begin{align}
	R =& \left( \dfrac{r - r_H}{r_H} \right)^{ - i \sigma} \left( \dfrac{r + r_H}{r_H} \right)^{i \sigma - \frac{3}{2}} {\rm e}^{q r} \nn \\
	&\qquad \times \sum_{ k=0 }^{ \infty } b_k \left( \dfrac{r - r_H}{r + r_H} \right)^k.\label{5.7}
\end{align}
Substituting the above equation into Eq.\eqref{3.1}, we obtain the following seven-term recurrence relation 
\begin{widetext}
	\begin{align}
		\alpha_0^r b_1 + \beta_0^r b_0 &= 0,  \\
		\alpha_1^r b_2 + \beta_1^r b_1 + \gamma_1^r b_0 &= 0,  \\
		\alpha_2^r b_3 + \beta_2^r b_2 + \gamma_2^r b_1 + \delta_2^r b_0 &= 0, \\
		\alpha_3^r b_4 + \beta_4^r b_3 + \gamma_4^r b_2 + \delta_4^r b_1 + \zeta_4^r b_0 &= 0, \\
		\alpha_4^r b_5 + \beta_5^r b_4 + \gamma_5^r b_3 + \delta_5^r b_2 + \zeta_5^r b_1 + \eta_5^r b_0 &= 0, \\
		\alpha_k^r b_{k+1} + \beta_k^r b_{k} + \gamma_k^r b_{k-1} + \delta_k^r b_{k-2} + \zeta_k^r b_{k-3} + \eta_k^r b_{k-4} + \kappa_k^r b_{k-5} &= 0, \quad (k=5,6,\cdots), \label{5.13}
	\end{align}
\end{widetext}
The recurrence relation \eqref{5.13} can be reduced by making a Gaussian elimination four times to a three-term recurrence relation
\begin{align}
	\tilde\alpha_k^r b_{k + 1} + \tilde\beta_k^r b_k + \tilde\gamma_k^r b_{k - 1} = 0, \qquad k = 2,3,\cdots.
\end{align}
The explicit procedure on how to get these coefficients with tilde can be found in Ref.\cite{Leaver:1990zz}. We do not show these coefficients in detail here since they have rather complicated forms. For a given set of the values of parameters $\{m,j,\lambda_{kjm},a,\mu\}$, the quasinormal frequency $\omega$ is a solution  the following continued fraction equation
\begin{equation}\label{5.15}
	\tilde\beta_0^r - {\tilde\alpha_0^r \tilde\gamma_1^r \over \tilde\beta_1^r - }
	{\tilde\alpha_1^r \tilde\gamma_2^r \over \tilde\beta_2^r - }
	{\tilde\alpha_2^r \tilde\gamma_3^r \over \tilde\beta_3^r - } \cdots = 0.
\end{equation}

In fact, the continued fraction equations \eqref{5.5} and \eqref{5.15} are two coupled algebraic equations for the unknown $\lambda_{kjm}$ and $\omega$. In order to numerically compute the QNMs, these two equations should be solved simultaneously.

\subsection{Results}\label{sec:5.2}

According to the method described above, we calculate the fundamental quasinormal frequencies for different values of the parameters $\left\lbrace k,j,m,a,\mu \right\rbrace$. To validate our code, we calculate the QNMs of massless scalar perturbation in a five-dimensional Schwarzschild BH and compare them with previous results in Ref.\cite{Zhidenko:2006rs}. 
We find agreement between our results and the ones in Ref.\cite{Zhidenko:2006rs}, which is shown in Table \ref{tab:2}.

%Table II
\begin{table}[tbph]
	\centering
	\caption{Comparison between our numerical results of the fundamental QNMs of massless scalar field on $5D$ MPBH  with $a=0$ and the results for $5D$ Schwarzschild BHs obtained in Ref.\cite{Zhidenko:2006rs}.}
	\label{tab:2}
	\renewcommand\arraystretch{1.25}
	\setlength{\tabcolsep}{2.5pt}{
		\begin{tabular}{c c c c c c c}
			\hline % horizontal line
			\hline
			$k$ & $j$ & $m$ & $\omega_{\rm Num}$ & $\omega_{\rm Sch}$ & \% {\rm Re} & \% {\rm Im} \\
			\hline
			$0$ & $0$ & $0$ & $0.54126 - 0.39585 i$ & $0.53384 - 0.38338 i$ & $1.37$ & $3.15$ \\
			$0$ & $0$ & $1$ & $1.01627 - 0.36352 i$ & $1.01602 - 0.36233 i$ & $0.02$ & $0.33$ \\
			$0$ & $1$ & $1$ & $1.51058 - 0.35776 i$ & $1.51057 - 0.35754 i$ & $0.00$ & $0.06$ \\
			\hline
			\hline
	\end{tabular}}
\end{table}

\begin{figure}[]
	\centering
	\includegraphics[scale=0.4]{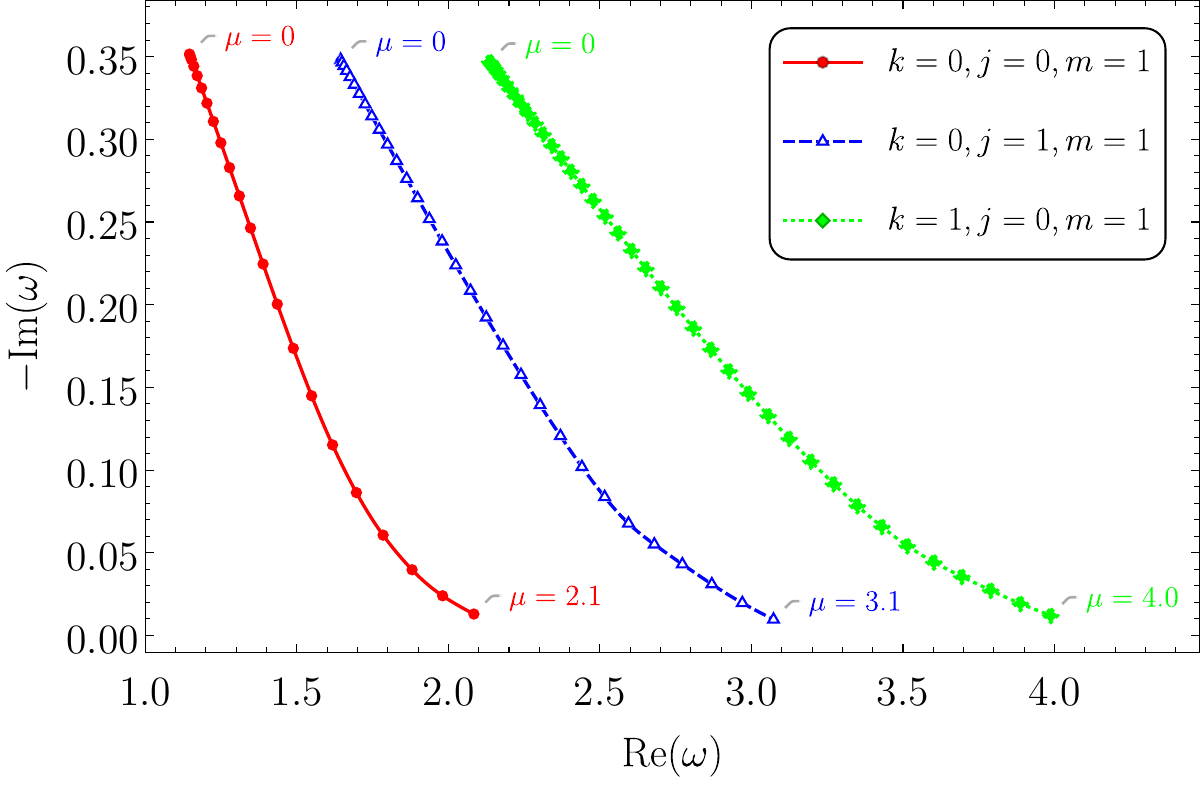}
	\caption{The fundamental scalar QNMs for different values of $\mu$. The points on each curve are plotted with the step of $\Delta \mu=0.1$, starting from $\mu=0$.
		The other parameters are chosen as $M=1,a=0.4$.}
	\label{fig:2}
\end{figure}

In Fig.\ref{fig:2}, the values of $\mathrm{Re}(\omega)$ and $-\mathrm{Im}(\omega)$ of the QNM frequencies are compared for different scalar field masses $\mu$ with step length 
$\Delta \mu = 0.1$. The red solid line, blue dashed line and green dotted line represent the $\{k=j=0,m=1\}$, $\{k=0, j=m=1\}$ and $\{k=m=1,j=0\}$ modes, respectively. 
Since $-\mathrm{Im}(\omega)>0$, all the QNMs are decaying modes and no instability appears.
It is easy to see that the imaginary parts of the QNMs tend to zero as the scalar mass $\mu$ increases. These long-living modes, called quasiresonances, are qualitatively 
the same as that found in four-dimensional Kerr BH cases \cite{Konoplya:2006br,Dolan:2007mj} and in Schwarzschild BH cases \cite{Ohashi:2004wr,Konoplya:2004wg,Zhidenko:2006rs}.
For larger quantum numbers, it is also found that the imaginary part $\mathrm{Im}(\omega)$  has a more slower tendency to zero. To calculate the cases with large enough 
scalar masses $\mu$, one needs a smaller step of $\Delta \mu$ and the continued fraction method should be improved by the Nollert technique \cite{Nollert:1993zz}.

\begin{figure}[]
	\centering
	\includegraphics[scale=0.4]{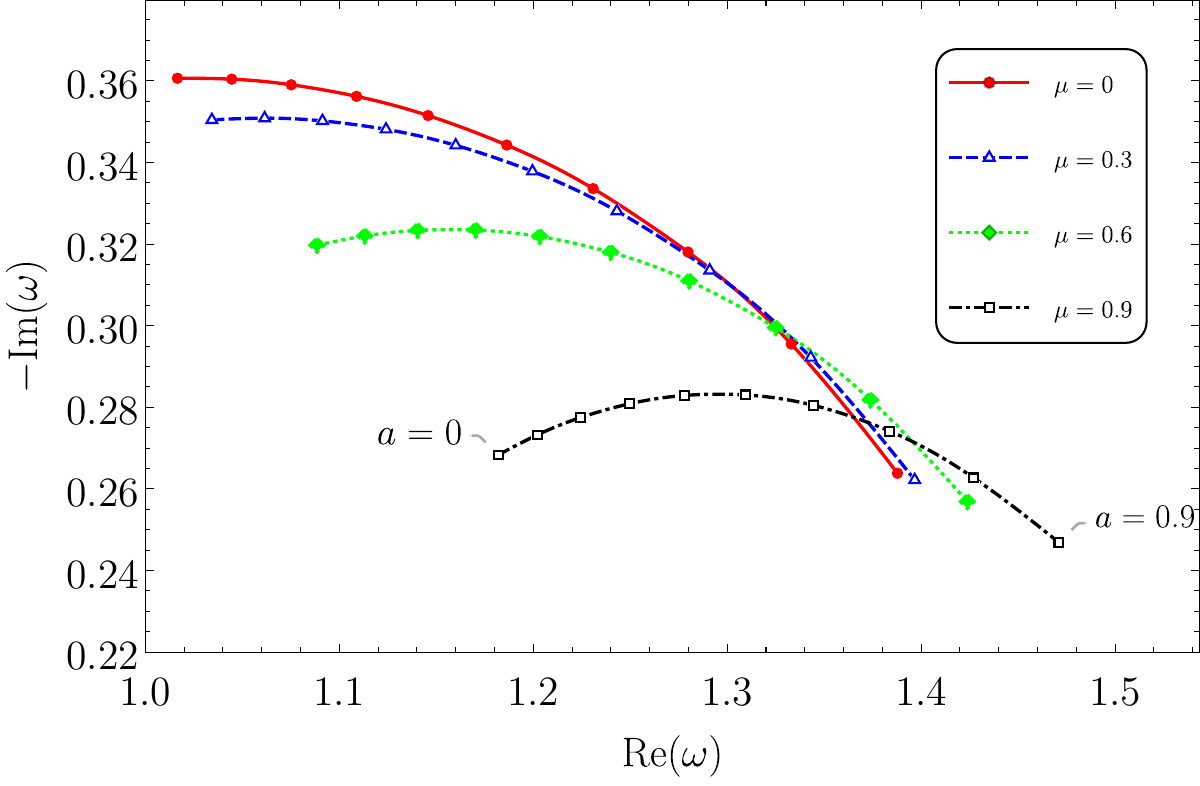}
	\caption{The fundamental QNM frequencies with quantum number $k=j=0,m=1$ are shown as a function of the rotation parameter $a$ of the MPBH. 
		Four curves correspond to different values of the scalar mass $\mu$.}
	\label{fig:3}
\end{figure}

In Fig.\ref{fig:3}, the dependence of the QNM frequencies of the lowest state ($\ell = m = 1$) on the rotation parameter $a$ is plotted.
The four curves correspond to the QNM modes with different scalar masses, $\mu=0, 0.3, 0.6$ and $0.9$. In each curve, it is plotted from left to right ten points
with $a=0, 0.1, 0.2, \cdots, 0.9$, respectively. First, we see that the QNMs of the massive scalar perturbation are all decaying modes and no instability appears.

\begin{figure*}
	\centering
	\begin{minipage}[]{0.48\textwidth}
		\vspace{0pt}
		\includegraphics[width=\textwidth]{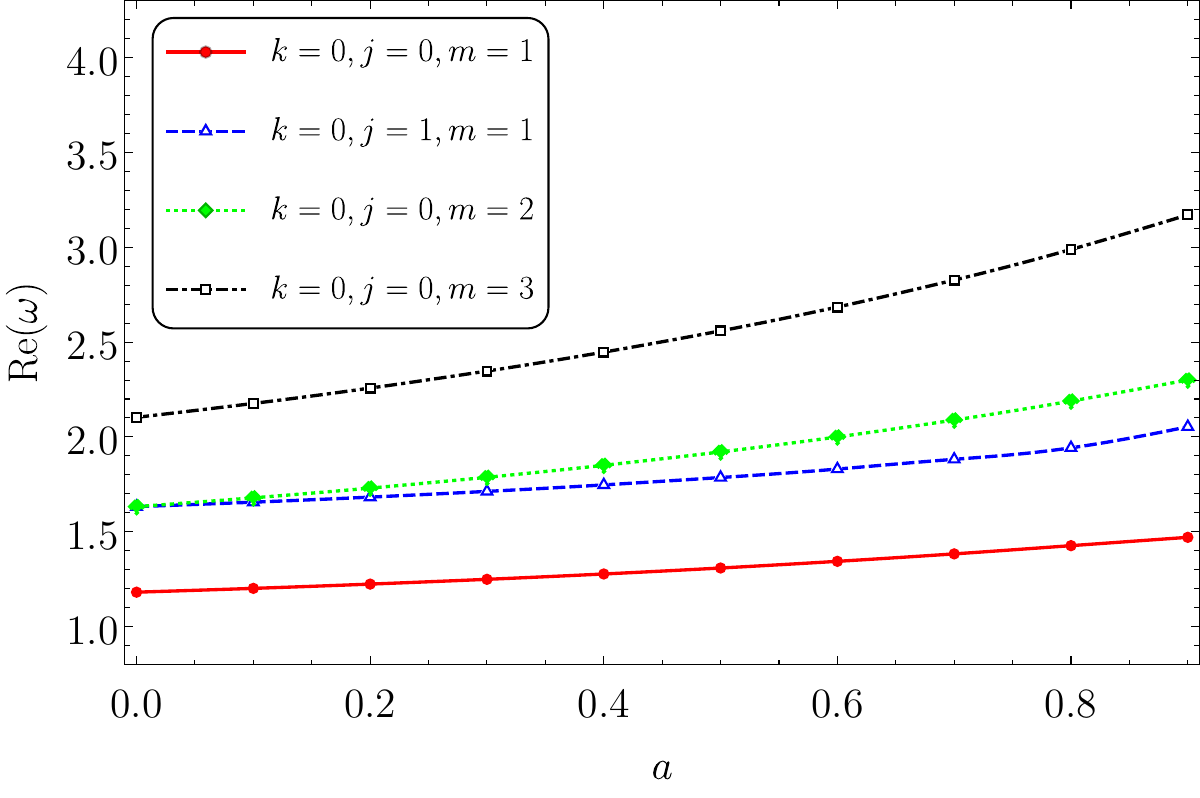}
	\end{minipage}
	\quad
	\begin{minipage}[]{0.48\textwidth}
		\vspace{0pt}
		\includegraphics[width=0.99\textwidth]{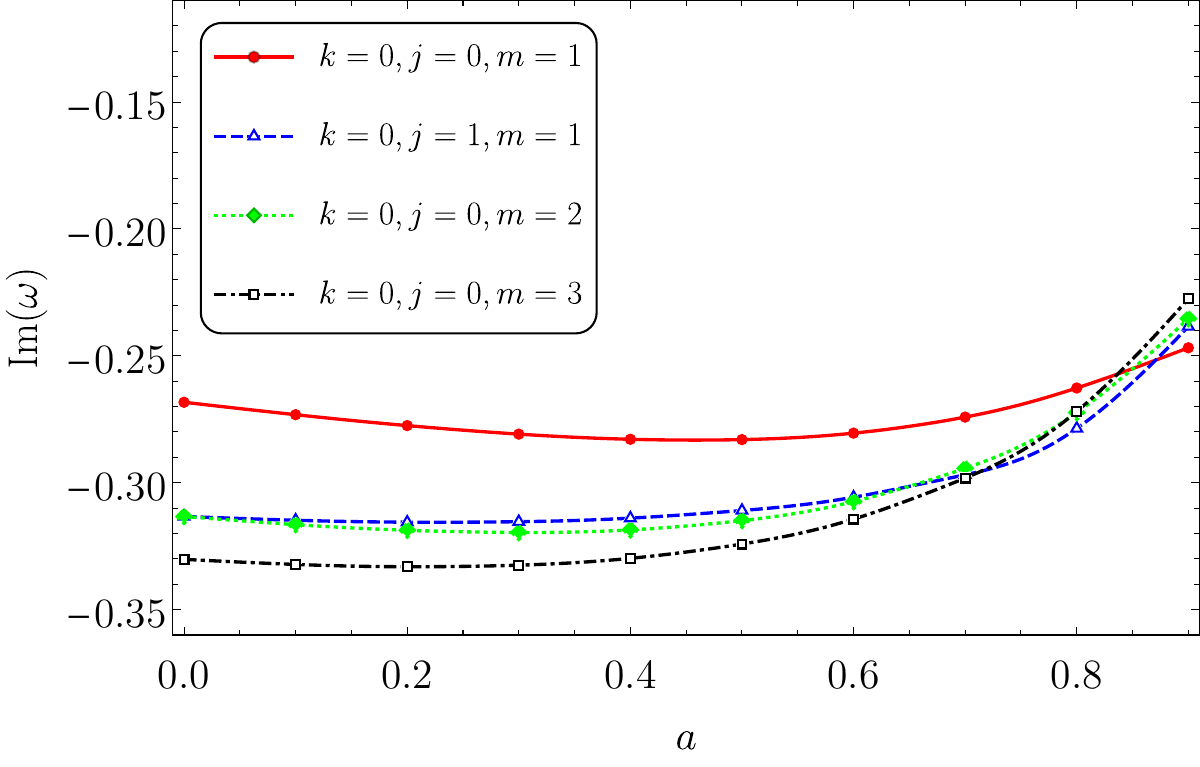}
	\end{minipage}
	\caption{The values of $\mathrm{Re}(\omega)$ and $\mathrm{Im}(\omega)$ of QNM frequencies against different values of the rotation parameter $a$. 
	Four sets of results with different quantum numbers $\left\lbrace \ell,j,m \right\rbrace$ are shown. The other parameters are chosen as $M=1,\mu=0.9$.}
	\label{fig:4}
\end{figure*}

Second, the damping rate, which is determined by $-\mathrm{Im}(\omega)$, is monotonically decreasing for massless QNMs as $a$ is increasing. However, the damping rate
of the QNMs of massive scalar perturbation is obviously non-monotonic with respect to $a$ when scalar mass $\mu$ is relatively large. 
It increases first and then decreases as $a$ increases, which is shown by the black 
curve in Fig.\ref{fig:3}. We also find that for a fixed set of quantum numbers $\left\lbrace \ell,m,j \right\rbrace$, this non-monotonic phenomenon becomes more and 
more obvious as $\mu$ increases. In contrast, for a fixed scalar mass $\mu$, the increasing of the quantum numbers suppresses this non-monotonic phenomenon, which is  
shown in Fig.\ref{fig:4}. 

Finally, the real parts of the QNMs of massless and massive scalar perturbation both monotonically increase as $a$ increases, which is also 
shown in Fig.\ref{fig:4}. 
It is noticeable that the dependence on the rotation parameter $a$ of the real parts and imaginary parts of the QNMs of massive scalar perturbation is
qualitatively the same as that in a $4D$ Kerr BH case \cite{Konoplya:2006br,Dolan:2007mj}. 

To find possible instability, we further calculate the fundamental QNM frequencies with another two sets of quantum numbers 
for different values of rotation parameter $a$ and scalar mass $\mu$.
In Table.\ref{tab:3}, the three quantum numbers are chosen as $j=0,k=m=1$, and  $k=j=m=1$ in Table.\ref{tab:4}. All found modes in the two tables 
are kept to the first six digits after the decimal point. It is easy to see that the imaginary parts of the QNMs are all negative, so they are all decaying modes and no unstable mode is found. 
\begin{table*}[]
	\centering
	\caption{The fundamental QNMs with $j = 0,k = m = 1$ for different values of $a$ and $\mu$.}
	\label{tab:3}
	\setlength{\tabcolsep}{3mm}
	\renewcommand\arraystretch{1.5}
	\begin{tabular}{c|c|c|c|c}
		\hline 
		\hline
		$a$ & $\mu=0$ & $\mu=0.3$ & $\mu=0.6$ & $\mu=0.9$ \\
		\hline
		$0.0$ & $2.007961 - 0.355712 i$ & $2.018538 - 0.352855 i$ & $2.050308 - 0.344314 i$ & $2.103385 - 0.330189i$ \\
		$0.1$ & $2.035047 - 0.355261 i$ & $2.045350 - 0.352508 i$ & $2.076303 - 0.344270 i$ & $2.128036 - 0.330621 i$ \\
		$0.2$ & $2.065353 - 0.353768 i$ & $2.075349 - 0.351140 i$ & $2.105385 - 0.343267 i$ & $2.155608 - 0.330193 i$ \\
		$0.3$ & $2.099136 - 0.351055 i$ & $2.108790 - 0.348572 i$ & $2.137805 - 0.341127 i$ & $2.186351 - 0.328733 i$ \\
		$0.4$ & $2.136696 - 0.346867 i$ & $2.145971 - 0.344553 i$ & $2.173861 - 0.337603 i$ & $2.220555 - 0.325998 i$ \\
		$0.5$ & $2.178379 - 0.340853 i$ & $2.187241 - 0.338731 i$ & $2.213895 - 0.332348 i$ & $2.258557 - 0.321654 i$ \\
		$0.6$ & $2.224575 - 0.332514 i$ & $2.232984 - 0.330612 i$ & $2.258290 - 0.324880 i$ & $2.300734 - 0.315236 i$ \\
		$0.7$ & $2.275690 - 0.321149 i$ & $2.283607 - 0.319502 i$ & $2.307446 - 0.314522 i$ & $2.347478 - 0.306104 i$ \\
		$0.8$ & $2.332014 - 0.305846 i$ & $2.339397 - 0.304499 i$ & $2.361642 - 0.300416 i$ & $2.399048 - 0.293467 i$ \\
		$0.9$ & $2.392369 - 0.286127 i$ & $2.399148 - 0.285159 i$ & $2.419587 - 0.282204 i$ & $2.453993 - 0.277111 i$ \\
		\hline
		\hline
	\end{tabular}
\end{table*}
\begin{table*}[]
	\centering
	\caption{The fundamental QNMs with $k = j = m = 1$ for different values of $a$ and $\mu$.}
	\label{tab:4}
	\setlength{\tabcolsep}{3mm}
	\renewcommand\arraystretch{1.5}
	\begin{tabular}{c|c|c|c|c}
		\hline 
		\hline
		$a$ & $\mu=0$ & $\mu=0.3$ & $\mu=0.6$ & $\mu=0.9$ \\
		\hline
		$0.0$ & $2.506329 - 0.354964 i$ & $2.514979 - 0.353107 i$ & $2.540942 - 0.347553 i$ & $2.584258 - 0.338349 i$ \\
		$0.1$ & $2.533676 - 0.354510 i$ & $2.542141 - 23352709 i$ & $2.567551 - 0.347321 i$ & $2.609954 - 0.338382 i$ \\
		$0.2$ & $2.565177 - 0.353047 i$ & $2.573425 - 0.351318 i$ & $2.598188 - 0.346140 i$ & $2.639524 - 0.337539 i$ \\
		$0.3$ & $2.601175 - 0.350423 i$ & $2.609174 - 0.348709 i$ & $2.633194 - 0.343858 i$ & $2.673306 - 0.335669 i$ \\
		$0.4$ & $2.642123 - 0.346411 i$ & $2.696021 - 0.339265 i$ & $2.673016 - 0.340252 i$ & $2.711739 - 0.332554 i$ \\
		$0.5$ & $2.688624 - 0.340684 i$ & $2.748526 - 0.331472 i$ & $2.718248 - 0.338002 i$ & $2.755404 - 0.327881 i$ \\
		$0.6$ & $2.741487 - 0.332752 i$ & $2.808427 - 0.320663 i$ & $2.769681 - 0.327625 i$ & $2.805070 - 0.321181 i$ \\
		$0.7$ & $2.801796 - 0.321780 i$ & $2.877230 - 0.304663 i$ & $2.828366 - 0.317298 i$ & $2.861745 - 0.311642 i$ \\
		$0.8$ & $2.871067 - 0.305600 i$ & $2.877230 - 0.304663 i$ & $2.895770 - 0.301832 i$ & $2.926841 - 0.297043 i$ \\
		$0.9$ & $2.963258 - 0.276453 i$ & $2.969035 - 0.275735 i$ & $2.986438 - 0.273555 i$ & $3.015686 - 0.269837 i$ \\
		\hline
		\hline
	\end{tabular}
\end{table*}

\section{Conclusions}

In this paper, we study the stability of a $5D$ singly rotating MPBH under massive scalar perturbations. 
It is found that a $5D$ singly rotating MPBH is stable against both the QBS modes and QNMs of the massive scalar perturbation.

We first consider the QBS modes which might lead to superradiant instability 
of the system through the black hole "bomb" mechanism. In the context of the perturbation theory of BH, we consider the constrains on the parameters of the system, which are
several important inequalities below Eqs.\eqref{4.1}\eqref{4.8}. Given these inequalities, 
we use an analytic method based on Descartes’ rule of signs to show that there is no potential well outside the event horizon of the MPBH. This means 
the $5D$ singly rotating MPBH is stable against the QBS modes of the massive scalar perturbations.

Then, we use Leaver's continued fraction method to numerically compute the QNMs of the massive scalar perturbation. We first 
introduce the specific steps of this method, and then make extensive calculation for the fundamental QNMs when the scalar mass $\mu$ is relatively small. We summarize our numerical results in several tables and figures. It is found that all the obtained fundamental QNMs are decaying modes, i.e. they are all stable.  

The damping rates of the QNMs are decreasing with the increasing of the scalar mass $\mu$. The fundamental
QNMs become quasiresonances with infinitely long life-time when the scalar mass becomes relatively large. 
The larger the scalar mass $\mu$ is, the more low-lying modes (modes with small quantum numbers) become the quasi-resonances.    
These properties are qualitatively the same as that found in other rotating BH cases \cite{Konoplya:2006br,Lu:2023par}. 

It is also found that the real parts of the fundamental QNMs monotonically increase with the increasing of the rotation parameter $a$, while 
the imaginary parts are not. However, the imaginary parts are always bounded within the region of negative values for different values of $a$.

\section*{Acknowledgments}
The authors are grateful to R. A. Konoplya and A. Zhidenko for their insightful comments. We also would like to thank Zhan-Feng Mai and Lihang Zhou for their valuable discussions. This work is partially supported by Guangdong Major Project of Basic and Applied Basic Research (No.2020B0301030008).

%reference
\providecommand{\href}[2]{#2}\begingroup\raggedright

\end{document}